\documentclass[journal]{IEEEtran}

  \ifCLASSINFOpdf

  \else

  \fi

\usepackage{multirow}
\usepackage{graphicx}
\usepackage{cite}
\usepackage{amsmath}
\usepackage{array}
\usepackage{amssymb}
\usepackage{algorithmicx}
\usepackage[ruled]{algorithm}
\usepackage{color}
\usepackage{microtype}
\usepackage{epstopdf}
\DeclareMathOperator*{\argmax}{arg\,max}
\DeclareMathOperator*{\argmin}{arg\,min}

\begin{document}
\title{\LARGE Blockwise Phase Rotation-Aided Analog Transmit Beamforming for 5G mmWave Systems}

\author{Md. Abdul Latif Sarker,  Igbafe Orikumhi, Dong Seog Han and Sunwoo Kim \vspace{-4ex}
\thanks{This work was supported by Institute for Information $\&$ communications Technology Promotion (IITP) grant funded by the Korea government (MSIT: 2016-0-00208, High Accurate Positioning Enabled MIMO Transmission and Network Technologies for Next 5G-V2X Services) (\textit{corresponding author: Sunwoo Kim})}
\thanks {M. A. L. Sarker, I. Orikumhi and S. Kim are with the Department of Electronics and Computer Engineering, Hanyang University, Seoul 04763, South Korea. (e-mail: abdul123@hanyang.ac.kr; oigbafe2@hanyang.ac.kr; remero@hanyang.ac.kr).}
\thanks{ D. S. Han are with the School of Electronic Engineering, Kyungpook National University, Daegu 41566, South Korea (e-mail: dshan@knu.ac.kr).}
\markboth{IEEE xxx xxxxx,~Vol.~xx, No.~xx, xx~2020}}
{}
\maketitle
\begin{abstract} 
 In this letter, we propose a blockwise phase rotation-aided analog transmit beamforming (BPR-ATB) scheme to improve the spectral efficiency and the bit-error-rate (BER) performance in millimeter wave (mmWave) communication systems. Due to the \textcolor{blue}{phase angle optimization issues} of the conventional analog beamforming, we design the BPR-ATB \textcolor{blue}{for reducing the rotated beamspace of the equivalent channel and improving the minimum Euclidean distance}. To verify the effectiveness of the proposed BPR-ATB scheme, we employ an Alamouti coding technique at the transmitter and evaluate the bit-error-rate performance for mmWave multiple-input and single-output systems. The simulation results show that the proposed BPR-ATB scheme outperforms the conventional discrete Fourier transform-based ATB scheme.
\end{abstract}
\IEEEpeerreviewmaketitle
\section{Introduction}
  
\IEEEPARstart{T}{he} millimeter-wave (mmWave) technology plays a major role  in the fifth-generation (5G) wireless communications owing to the large bandwidth \cite{6484896}  and spectral efficiency  \cite{ 8371237, 8964409}. 
The mmWave technology operates in the 30 to 300 GHz band \cite{6484896, 8371237}, hence the large spectral resource in contrast with  microwave technologies operating in the sub 6 GHz band \cite{1369651}.
 Typically, mmWave system requires  massive antenna arrays, which are equipped with the base station (BS) for achieving a highly directive beamforming \cite{7400949}. For deployment of this system, the leading barriers are the hardware limitations, the channel sparsity, the free-space path loss, beamforming construction, and phase angle optimization.\par
 The sparse nature of the channel and the discrete Fourier transform-based analog beamforming (DFT-ATB) schemes have been investigated  in \cite{6484896, 8964409, 7400949, 8777168, 8401880, 6928432,8565897}. The authors designed a joint antenna selection based transmit beamforming in \cite{8401880}. A phase control DFT based hybrid precoding scheme is presented in \cite{6928432, 8565897}. Particularly, the traditional analog beamforming incurs a quantization error in communication systems owing \textcolor{blue}{to their low minimum Euclidean distance \cite{6928432, 1369651, 6378483}}. In addition, the conventional DFT-ATB scheme shows a `beam squint' challenge with a wideband channel \cite{8777168, 6692283}. The `beam squint' leads a higher channel spreading factor due to the structural leakage of the conventional analog beamforming. To get a better\textcolor{blue}{ minimum Euclidean distance of analog precoding}, the authors proposed a Golden-Hadamard (GH) based precoding in \textcolor{blue}{\cite{8428615}}. Although the GH scheme achieved a remarkable bit-error-rate (BER) performance in microwave systems, the scheme shows a phase angle optimization problem in highly directive wireless systems due to their wide rotated beamspace.  Hence, we design a blockwise phase rotation-aided analog transmit beamforming (BPR-ATB) for mmWave communication systems.\par

In this article, we propose a BPR-ATB scheme \textcolor{blue}{to minimize the rotated  beamspace of the equivalent channel and improving the minimum Euclidean distance} of  traditional analog beamforming such as the DFT-ATB scheme. To this end, we seek \textcolor{blue}{to obtain an efficient rotated beamspace of the equivalent channel and improve the minimum Euclidean distance}. We first run back \textcolor{blue}{\cite[eq. (9)]{8428615}} and then design a BPR-ATB scheme to get the effective rotated beamspace and generate a satisfactory spectral efficiency of the mmWave communications. After that, we implement the proposed BPR-ATB scheme with Alamouti code and set a power factor-based parameter $\kappa$ in the BER performance metric. Finally, we show the superiority of the proposed BPR-AB scheme over the DFT-ATB scheme in terms of a downlink mmWave multiple-input and single-output (MISO) systems through computer simulations.
\raggedbottom
\section{Channel and Signal Models}
 We consider a downlink mmWave MISO system with $N_t$ transmit antennas and  a single antenna receiver. Then the received signal vector $\mathbf{y}\in\mathbb{C}^{1\times T}$ can be modeled as
\begin{equation}
\textcolor{blue}{\mathbf{y}=\sqrt{\frac{PN_t}{L}}\mathbf{h}^H\mathbf{X}+\mathbf{z}},\label{eq:received signal}
 \end{equation}
 where $P$ denotes the transmit power, $L$ is the number of paths, $\mathbf{X}$ is the $N_t\times T$ space-time codeword matrix, $T$ is the number of time slot, and \textcolor{blue}{$\mathbf{z}\sim\mathcal{CN}(0,1)$ is additive white Gaussian noise vector with zero-mean and unit variance. The narrow-band mmWave channel $\mathbf{h}\in \mathbb{C}^{N_t\times 1}$ with $L$ propagation paths \cite{6484896, 7400949}, that is
  \begin{equation}
\mathbf{h}=\sum_{l=1}^{L}\alpha_l\mathbf{a}(\theta_l),
 \end{equation}}
where $\alpha_l$ is the complex gain of the $l$-th path, $\theta_l$ represents the angle of departure of the $l$-th path, $\mathbf{a}(\theta_l)$ denotes the transmit steering vector of the $l$-th path, which is given by
\begin{equation}
\mathbf{a}(\theta_l)=[1,e^{j\frac{2\pi d}{\lambda}\sin\theta_l},...,e^{j\frac{(N_t-1)\pi d}{\lambda}\sin\theta_l}]^T,
\end{equation}
 the wavelength, $\lambda=c/{f_{c}}$, $c$ is the speed of light, $f_{c}$ is the carrier frequency, and \textcolor{blue}{$d=\lambda/2$} is the antenna spacing.
 \raggedbottom
\section{Blockwise Phase Rotation-Aided Analog Transmit Beamforming (BPR-ATB) Scheme}
\subsection{BPR-ATB matrix design} Due to \textcolor{blue}{the phase angle optimization and beam squint issues in the high dimensional ATB scheme, we first run back \cite[eq. (9)]{8428615}} and then we design the Golden Hadamard based BPR-ATB scheme in this section.
\begin{figure}[b!]
	\centering{}\includegraphics[width=4in,height=4in,keepaspectratio]{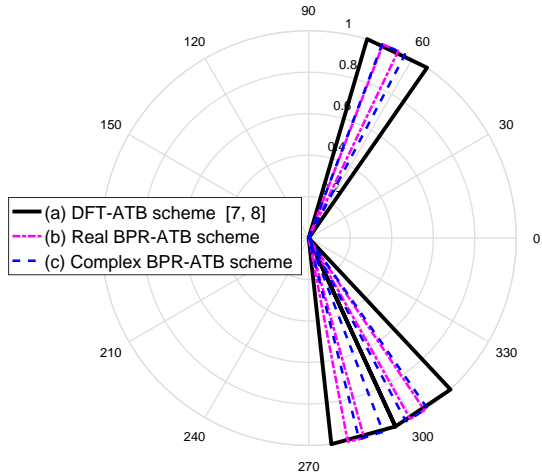}\caption{\textcolor{blue}{The rose diagram of the rotated beamspace for the effective equivalent mmWave channel (a) the conventional DFT-ATB scheme \cite{8401880,6928432} in solid lines, (b) and (c) the effective equivalent mmWave channel with the proposed BPR-ATB scheme in doted lines.}}
\end{figure}
Let \textcolor{blue}{the number of total transmit antennas $N_t=2^q$ and $\xi=n\{(1+n)^q-(1-n)^q\}/2^q$ where $q=\textrm{log}_2N_t$ and $n$ denotes the root of the geometric number. Consider the space-time codeword matrix $\mathbf{X}$ as
\begin{equation}
\mathbf{X}=\mathbf{F}_{2^q}\mathbf{S},\label{eq: space-time}
\end{equation}
where $\mathbf{S}$ is a $2^{q-1}\times T$ orthogonal space time block code matrix, $\mathbf{F}_{2^q}$ be a propose $2^{q}\times 2^{q-1}$ BPR-ATB matrix constructed by $2^{q-1}$ columns of a $2^{q}\times 2^{q}$ recursive Golden-Hadamard matrices as follows
\begin{equation}
 \mathbf{F}_{{2^q}}=\frac{g}{\sqrt{\xi}}\left[\begin{IEEEeqnarraybox*}[][c]{,c/c/c/c,}
  \mathbf{W}_{2^{q-1}}\mathbf{A}_{2^{q-1}}&\mathbf{W}_{2^{q-1}}\mathbf{B}_{2^{q-1}}\\
  \mathbf{W}_{2^{q-1}}\mathbf{B}_{2^{q-1}}& -\mathbf{W}_{2^{q-1}}\mathbf{A}_{2^{q-1}}
  \end{IEEEeqnarraybox*}\right],
\end{equation}
 where $g$ denotes the golden number \cite{8428615, Olsen2006}, $\mathbf{W}_{2^{q-1}}$ is a $2^{q-1}\times 2^{q-1}$ block Hadamard matrix, 
  $\mathbf{A}_{2^{q-1}}=\textrm{diag}\{e^{j\phi_{\nu_1}}\}$ and  $\mathbf{B}_{2^{q-1}}=\textrm{diag}\{e^{j\phi_{\nu_2}}\}$ are the $2^{q-1}\times 2^{q-1}$ block-diagonal phase rotation matrix, $\nu_1$ and $\nu_2$ are the block-order of  $\mathbf{A}_{2^{q-1}}$ and $\mathbf{B}_{2^{q-1}}$.}
By substituting \eqref{eq: space-time} in \eqref{eq:received signal}, the system can achieve a spectral efficiency for MISO system given as \begin{equation}
R=\log_{2}\left\{1+\frac{P}{\sigma^2}\mathbf{h}^H\mathbf{F}_{2^q}\mathbf{F}_{2^q}^{H}\mathbf{h}\right\}.
\end{equation}
\subsection{Problem formulation} Particularly, the phase rotation on the transmitted signals is effectively equivalent to rotating the phases of the corresponding channel coefficients. It should be noted that, the conventional DFT-ATB scheme generates a satisfactory array gain with equivalent channel \cite{8401880, 6928432}, but this scheme suffers a phase angle optimization problem, \textcolor{blue}{which leads to a wide beamspace of the equivalent channel as shown in Fig. 1. Using by (6) and \cite{8401880}, the optimization problem can be formulated 
 \begin{equation}
 \begin{split}
\mathbf{f}_{2^q}^{opt}&=\argmax_{\mathbf{f}_{2^q}}R\\
& s.t.\hspace{3pt}\mathbf{f}_{2^q}\in \left\{\mathcal{F},0\right\}^{2^q},\\
& \textrm{and} \hspace{5pt}||\mathbf{f}_{2^q}||_0=2^{q-1},
\end{split}
\end{equation}
\raggedbottom
where $\mathbf{f}_{2^q}$ denotes the vector of $\mathbf{F}_{2^q}$. We observe that the (7) maximizes the spectral efficiency but it has a non-convex objective function, which conducts a phase angle optimization problem. In addition, the spectral efficiency (6) is a correctly monotone enhancing function of beamforming gain $|\mathbf{h}\mathbf{f}_{2^q}|$. To simplify this problem, we formulate the phase angle optimization as below:}

Let $h_{\nu}$ be the $\nu$-th element of $\mathbf{h}$ and $\mathbf{f}_{2^q,\nu}\in\mathcal{F}$ be the effective analog beamforming vector. Thus, the phase angle optimization problem is given by
 \begin{equation}
 \begin{split}
\varphi_{\nu}^{opt}&=\argmax_{\varphi_{\nu}}\left|\sum_{\nu=1}^{2^q} h_{\nu}^{*}\mathbf{f}_{2^q,\nu}(\varphi_{\nu})\right|\\
& s.t.\hspace{3pt}\varphi_{\nu}\in \left\{\frac{2\pi b}{2^q} \bigg|b=0,1...,2^q-1\right\},\\
& \textrm{and} \hspace{5pt}\nu=1,...,2^q,
\end{split}
\end{equation}
\textcolor{blue}{where $\mathbf{f}_{\nu}(\varphi_{\nu})=e^{j\varphi_{\nu}}/\sqrt{2^q}$}. We observe the \textcolor{blue}{(8)} is still leading an optimization problem due to the global phase angle, which generates an extensive beamspsce with a mmWave channel. As a result, the user suffers from a high computational burden to  optimize the global phase angle.
To overcome the phase angle optimization problem, we reformulate \textcolor{blue}{(8)} and propose the BPR-ATB based algorithm in the Subsection C.
\subsection{Proposed BPR-ATB based algorithm}
 \textcolor{blue}{Let $\varphi_{\nu}\in\{{\phi_{\nu_1},\phi_{\nu_2}}\}$, where $\phi_{\nu_1}$ and $\phi_{\nu_2}$ are the block phase angle of $\mathbf{A}_{2^{q-1}}$ and $\mathbf{B}_{2^{q-1}}$. We set $\phi_{\nu_1}$ and $\phi_{\nu_2}$} in the designed transmit beamformer of $\mathbf{F}_{2^q}$. Then the optimal block phase angle is given by
 \textcolor{blue}{\begin{equation}
  \begin{split}
 \varphi_{\nu}(\phi_{\nu_1}^{opt},\phi_{\nu_2}^{opt})&=\argmax_{\phi_{\nu_1},\phi_{\nu_2}}\left|\sum_{\nu=1}^{2^{q}}h_{\nu}^{*}e^{j\phi_{\nu_1}}e^{j\phi_{\nu_2}}\right|\\
 & s.t.\hspace{3pt}\phi_{\nu_1}\in \left\{\frac{2\pi b_1}{2^{q-1}}\bigg|b_1=0,1...,2^{q-1}-1\right\},\\
 & \hspace{3pt}\phi_{\nu_2}\in \left\{\frac{2\pi b_2}{2^{q-1}}\bigg|b_2=2^{q-1}...,2^{q}-1\right\},\\
 & \nu_1=\nu_2 \hspace{5pt} \textrm{and} \hspace{5pt} b_1\neq b_2.
 \end{split}
 \end{equation}}
 \begin{algorithm}[t!]
\caption{Proposed BPR-ATB based Algorithm}
\begin{algorithmic}[1]
 \State	\textcolor{blue}{ \textbf{Input parameters:} $\mathbf{h}$, $b_1$, $b_2$ $q$}.
	\medskip
	\State \textcolor{blue}{\textbf{Output:} $\mathbf{f}_{2^q, v}^{opt}$}.
	\medskip
	\State \textcolor{blue}{\text{Obtain} $ \varphi_{\nu}(\phi_{\nu_1}^{opt}, \phi_{\nu_2}^{opt})$ }.
	\medskip
	\State \textbf{Begin} $\mathcal{C}:=\{\nu\}$, $\nu\in\{\mathcal{S}_{\nu_1},\mathcal{S}_{\nu_2}\}$, and $\mathcal{S}_{\nu_1}=\mathcal{S}_{\nu_2}=\emptyset$. 
	\medskip
	\State \textbf{for} $\nu_1=1:2^{q-1}$ \textbf{do}
	\State \text{Find} \par
	${\nu}^{opt_1}=\argmax_{\nu\in{\mathcal{C}}}\left|\sum_{\nu_1\in\mathcal{S}_{\nu_1}}(h_{\nu_1}^{*}+h_{\nu}^{*}e^{j\phi_{\nu_2}})e^{j\phi_{\nu_1}}\right|$.
	\State$\mathcal{S}_{\nu_1}$ $:=\mathcal{S}_{\nu_1}\bigcup{\{\nu}^{opt_1}\}$.
	\State $\mathcal{C}$ $:=\mathcal{C}\hspace{2pt}\text{mod}\hspace{2pt}\{{\nu}^{opt_1}\}$.
    \State \textbf{end for}
	\State \textbf{for} $\nu_2=1:2^{q-1}$ \textbf{do}
	\State \text{Find} \par
	${\nu}^{opt_2}=\argmax_{\nu\in{\mathcal{C}}}\left|\sum_{\nu_2\in\mathcal{S}_{\nu_2}}(h_{\nu_2}^{*}+h_{\nu}^{*}e^{j\phi_{\nu_1}})e^{j\phi_{\nu_2}}\right|$.
	\State$\mathcal{S}_{\nu_2}$ $:=\mathcal{S}_{\nu_2}\bigcup{\{\nu}^{opt_2}\}$.
	\medskip
	\State $\mathcal{C}$ $:=\mathcal{C}\hspace{2pt}\text{mod}\hspace{2pt}\{{\nu}^{opt_2}\}$.
	\State \textbf{end for}
	\medskip
	\State \textcolor{blue}{\text{Obtain} $\mathbf{f}_{2^q, v}^{opt}=(ge^{j\phi_{\nu_1}^{opt}}e^{j\phi_{\nu_2}^{opt}})/{\sqrt{\zeta}}$ \text{according to} (9).}
	\end{algorithmic}
    \end{algorithm}
Consider $\mathcal{C}$ as the set of indexes of useful antennas and $\nu\in\{\mathcal{S}_{\nu_1},\mathcal{S}_{\nu_2}\}$, where $\mathcal{S}_{\nu_1}$ and $\mathcal{S}_{\nu_2}$ are the subset of beam indices. Based on (9), we demonstrate the proposed BPR-ATB scheme in \textbf{Algorithm 1}.
 \section{Simulation Results and Discussion}
In this section, we  compare the proposed BPR-ATB scheme against the conventional DFT-ATB scheme via computer simulations. To show the superiority of the proposed scheme, we employ a complex Alamouti coding technique at the transmitter. For example, if we use the $k$-th entry of a complex alamouti code with time slot $T=2$ in (4) where we consider $T$ is equal to the number of radio frequency chain, then the codeword matrix $\mathbf{X}$ is given by 
\begin{eqnarray}
\begin{split}
\mathbf{X}&=\sqrt{\gamma_0\kappa}\mathbf{F}\mathbf{S}_{k}\\
&=\sqrt{\gamma_0\kappa}\mathbf{F}\left[\begin{IEEEeqnarraybox*}[][c]{,c/c/c,}
  s_{11}&-s_{21}^{*}\\
  s_{21}& s_{11}^{*}
\end{IEEEeqnarraybox*}\right],
\end{split}
\end{eqnarray}
where $\mathbf{S}_k$ is the $k$-th entry of $2\times 2$ complex Alamouti code \cite{730453}, symbols $s_{11}$ and $s_{21}$ belongs to a quadrature amplitude modulation (QAM) constellation, $\kappa=|\mathbf{F}_{i,j}|^2$ is a power factor of the $(i,j)$ element of analog beamforming $\mathbf{F}$, and $\gamma_0$ is the received signal-to-noise power (SNR). The $\kappa$ value depends on the structure of $\mathbf{F}$ matrix (see in Table I).\par

\begin{table}[b!]
\centering
\caption{\small Different power factor of RF precoding}
\begin{tabular}{|*{9}{p{8.47cm}|}}
\hline
\centering
\small Power factor of RF precoding, $\kappa=|\textbf{F}_{i,j}|^2$
\end{tabular}\\
\begin{tabular}{|c|c|c|c|}
\hline
\multirow{2}*[-.1ex]{\small DFT\cite{1369651, 6928432}}&\multirow{2}*[-.1ex]{\small HA\cite{6468994}} &\multicolumn{2}{c|}{\small BPR}\\ \cline{3-4}
 & & $g$-$\textrm{real value case}$ & $g$-$\textrm{complex value case}$ \\ \hline
$\cfrac{{1}}{{N_t}}$ & $\cfrac{{1}}{{N_t}}$ & $\cfrac{(1+\sqrt{5})^2}{4{\xi}}$ & $\cfrac{(j+\sqrt{3})^2}{4{\xi}}$\\[1.8ex] \hline  
\end{tabular}\\
\end{table}

Consider $\mathbf{S}_{k}$ and $\mathbf{S}_{l}$ as the transmitted and decoded space-time codewords, respectively, \text{where} $k\neq l$. The union bound on the bit-error-rate (BER) is formulated as \cite{Proakis2007} 
 \begin{equation}
 BER\leq\sum_{k\neq l} \frac{e(\mathbf{S}_{k}, \mathbf{S}_{l})}{\log_{2}(M)}Q\left(\Xi_{k,l}\sqrt\frac{\gamma_0\kappa}{2}\right),
 \end{equation}
where $M$ denotes the constellation size, the operator $Q(\cdot)$ represents the Q-function \cite{Proakis2007}, $\Xi_{k,l}=\|\mathbf{h}_{eq}^{H}\mathbf{e}_{k,l}\|_{F}$ where the operator $\|\cdot\|_{F}$ denotes a Frobenius norm,  $\mathbf{e}_{k,l}=\mathbf{S}_{k}-\mathbf{S}_{l}$ represents an error matrix between the codewords $\mathbf{S}_{k}$ and $\mathbf{S}_{l}$, $e(\mathbf{S}_{k}, \mathbf{S}_{l})$ is the Hamming distance between the bit mappings corresponding to the vectors $\mathbf{S}_{k}$ and $\mathbf{S}_{l}$. Invoking the Chernoff upper bound $Q(x)\leq e^{-x^2/2}$ in (11), the pairwise error probability can be upper bounded with an equivalent channel $\mathbf{h}_{eq}$ as
\begin{equation}
\textrm{P}_{r}(\mathbf{S}_{k}\rightarrow\mathbf{S}_{l}|\mathbf{h}_{eq})\leq e^{-\frac{\gamma_0\kappa\Xi_{k,l}^{2}}{4}},
\end{equation}
where $\mathbf{h}_{eq}=\mathbf{F}^{H}\mathbf{h}$, \textcolor{blue}{ and $\Xi_{k,l}^{2}$ is the minimum Euclidean distance as
\begin{equation}
\Xi_{k,l}^{2}= \argmin_{\mathbf{S}_{k}\neq \mathbf{S}_{l}}||\mathbf{h}^{H}\mathbf{F}_{2^q}\mathbf{e}_{k,l}||_{F},
\end{equation}
where the property of the error matrix $\mathbf{e}_{k,l}\mathbf{e}_{k,l}^{H}=a\mathbf{I}$} for the orthogonal space-time block coding, $a$ is a constant depending on the constellation [see in {A\small{PPENDIX}} A].
\begin{table}[b!]
\centering
\caption{\small Simulation Parameters}
 \begin{tabular}{|c|c|c|c|c|c|c|} 
\hline
\small Total number of transmit antennas & $N_{t}=4$ \\
\small Total number of receive antenna & $N_{r}=1$ \\
\small Number of time slot & $T=2$ \\
\small Channel path & $L=3$ \\
\small SNR & $\gamma_0=20$ dB \\
\small Carrier frequency & $f_c=60$ GHz \\
\small Wavelength & $\lambda=5$ mm \\
\small Antenna spacing distance & $d=\lambda/2$ \\
\small  Modulation scheme & 64 QAM\\\hline
\end{tabular}\\
\end{table}

\begin{figure}[b!]
	\centering{}\includegraphics[width=3.3in,height=3.3in,keepaspectratio]{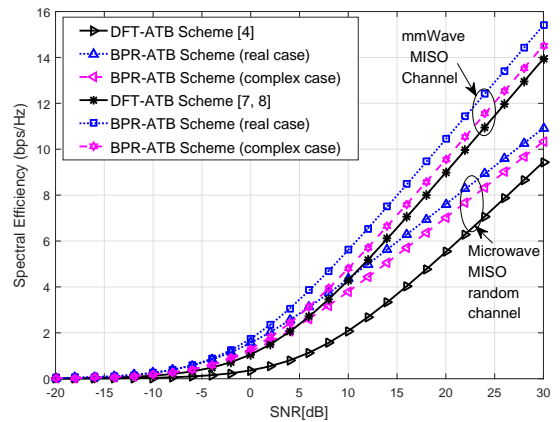}\caption{Spectral Efficiency versus SNR ($N_t=4, N_{RF}=T=2$, and $N_r=1$  with different $\kappa$ values of the analog transmit beamforming.}
\end{figure}

\begin{figure}[t!]
	\centering{}\includegraphics[width=3.3in,height=3.3in,keepaspectratio]{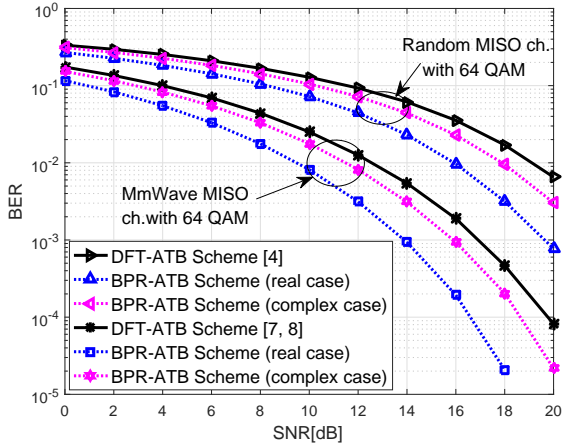}\caption{BER performance of 64QAM modulated Alamouti coding with different $\kappa$ values of analog transmit beamforming.}
\end{figure}


Throughout the simulations, we assumed the parameters of Table II with different $\kappa$ factor of Table I. \textcolor {blue}{In the case of mmWave system, we use a sparse geometric mmWave channel model \cite{7400949} with $L=3$ paths and angles of departure uniformly distributed over $[-\pi/2,\pi/2]$. Since the exhaustive search method (8) obviously measures a high complexity, as a results, we designed a sub-optimal BPR-ATB algorithm in \textbf{Algorithm 1}. The total time complexity of the \textbf{Algorithm 1} is $\mathcal{O}(2^q\zeta)$, which is adoptable even if $q$ is large and $2^q<\zeta$. We further consider the $2\times 2$ block-diagonal rotation matrices with block phase angle $\phi_{\nu_1}=2\pi b_1/2^{q-1}$ and $\phi_{\nu_2}=2\pi b_2/2^{q-1}$ when $b_1\in\{0,1\}$ and $b_2\in\{2,3\}$ is applied in (9)}. We compute and set the parameter $\kappa$ in the simulations such as $\kappa=0.25$ for both the DFT-ATB scheme \cite{1369651} and the Hadamard based ATB scheme \cite{6468994}, \textcolor {blue}{$\kappa=0.52$ for the proposed real BRP-ATB scheme, and $\kappa=0.33$ for the proposed complex BRP-ATB scheme} while $q=2$ is applied in Table I.\par
Fig. 2 illustrates the spectral efficiency comparisons of the proposed BPR-ATB scheme against the conventional DFT-ATB scheme. 
For the mmWave setup, the performance of the spectral efficiency is around $1.8$ bits/s/Hz at $30$ dB SNR values. The proposed BPR-ATB scheme also achieved good spectral performance with the traditional MISO random channel environment as shown in Fig. 2.

Fig. 3 shows the BER performance of the $4\times 2$ BPR-ATB scheme using $2\times 2$ Alamouti coding and 64QAM constellations. We see that the DFT-ATB scheme achieves a worse spectral efficiency and BER performance in \cite{1369651, 8401880, 6928432} for both mmWave and the familiar microwave MISO systems owing to their rotated wide beamspace (see in  Fig.1) and a teeny $\kappa$ factor (see in Table I). The proposed BPR-ATB scheme provides at least $2$ dB more BER performance with the DFT-ATB scheme. Furthermore, both spectral efficiency and BER performance of a complex case of the proposed BPR-ATB scheme is slightly worsened than that of a real case of the proposed BPR-ATB scheme because of the reduction of the value of the Golden ratio.
\section{Conclusions}
We proposed a BPR-ATB scheme to minimize the rotated beamspace of the equivalent mmWave  channel and improve the error performance of the systems. We verify  the effectiveness of the proposed BPR-ATB scheme by computer simulation and compare the performance with the conventional DFT-ATB scheme. The traditional DFT-ATB scheme exhibited a worse spectral efficiency about $1.8$ bits/s/Hz and BER performance difference of $2$ dB when compared with the proposed BPR-ATB scheme. 
Hence, the proposed BPR-ATB scheme can be extended further to the next generation multiple-input and multiple-output non-orthogonal multiple-access (MIMO-NOMA) systems, which will be explored in future studies.

\appendices \section{}
From (11), we can measure as a general single-input and single-output case of BER as follows:
 \begin{equation}
  \begin{split}
P_{e,b}&=\int_{0}^{\infty}Q\left(a\sqrt\gamma\right)p_{\gamma}(\gamma)d\gamma\\
&=\frac{1}{\pi}\int_{0}^{\frac{\pi}{2}}\mathcal{M}_{\gamma}\left(\frac{-a^2}{2\textrm{sin}^2\theta}\right)d\theta
\end{split}
\end{equation}
where $\gamma=\kappa\gamma_0/2$. The Gaussian Q-fuction $Q(x)$ is given by \cite{Proakis2007}
\begin{equation}
Q(x)=\frac{1}{\pi}\int_{0}^{\frac{\pi}{2}}\textrm{exp}\left(\frac{-x^2}{2\textrm{sin}^2\theta}\right)d\theta.
\end{equation}
It is noted that in (14), $\mathcal{M}_{X}\left(-s\right)=\int_{0}^{\infty}e^{-sx}p_{X}(x)dx$ is a moment generating function (MGF) of random variable $X$ where $\mathcal{M}_{\gamma}\left(s\right)=\int_{0}^{\infty}e^{s\gamma}p_{\gamma}(\gamma)d\gamma$ is the Laplace transform of $p_{\gamma}(\gamma)$ with the exponent reversed sign. By using the Rayleigh channel, we can consider $p_{\gamma}(\gamma)=\textrm{exp}(-\gamma/\overline{\gamma})/{\overline{\gamma}}$, where $\gamma\geq 0$ and $\overline{\gamma}$ idenotes the average SNR per bit. Hence, the Laplace transform is given by
\begin{equation}
\mathcal{M}_{X}\left(-s\right)=\frac{1}{1+s\overline{\gamma}}, \hspace{3pt} s>0.
\end{equation}
Substituting (16) into (14) gives
\begin{equation}
P_{e,b}(a,\overline{\gamma})=\frac{1}{2}-\frac{a}{2}\sqrt\frac{\overline{\gamma}}{2+a^2\overline{\gamma}}.
\end{equation}
 For the simplicity, (17) is derived for the binary phase shift keying (when $a=1$) modulation scheme, which can be straightforwardly extended to other modulation cases. For example, M-ary phase shift keying modulation scheme: we consider $a^2=2\textrm{sin}^{2}(\pi/M)$ and applying (16) in (14), then we follow directly (17) and obtain the bit-error-rate probability for Rayleigh fading channel as
\begin{equation}
\begin{split}
P_{e,b}(a,\overline{\gamma}, M)&=\frac{(M-1)}{M}-\frac{\sqrt\mu}{2}+\\
&\frac{(M-1)\sqrt\mu}{M}\textrm{tan}^{-1}\left(\sqrt\mu \textrm{cot}\frac{\pi}{M}\right),
\end{split}
\end{equation}
where $\mu=(\overline{\gamma}\textrm{sin}^{2}{\frac{\pi}{M}})/({1+\overline{\gamma}\textrm{sin}^{2}{\frac{\pi}{M}}}).$  Similarly, using (14) and (16), we can measure the BER probability for M-ary QAM modulation as
\begin{equation}
\begin{split}
P_{e,b}(a,\overline{\gamma}, M)&=\frac{4\zeta}{\pi}\int_{0}^{\frac{\pi}{2}}\left(1+\frac{3\overline{\gamma}}{2(M-1)\textrm{sin}^2\theta}\right)^{-1}d\theta\\
&-\frac{4\zeta^{2}}{\pi}\int_{0}^{\frac{\pi}{4}}\left(1+\frac{3\overline{\gamma}}{2(M-1)\textrm{sin}^2\theta}\right)^{-1}d\theta,
\end{split}
\end{equation}
where $a^2=3/(M-1)$ and $\zeta=1-(1/\sqrt{M})$.
 \ifCLASSOPTIONcaptionsoff
  \raggedbottom
    \newpage
  \fi
\bibliographystyle{IEEEtran}  
\bibliography{IEEEabrv,Reference}

\begin{thebibliography}{10}
\providecommand{\url}[1]{#1}
\csname url@samestyle\endcsname
\providecommand{\newblock}{\relax}
\providecommand{\bibinfo}[2]{#2}
\providecommand{\BIBentrySTDinterwordspacing}{\spaceskip=0pt\relax}
\providecommand{\BIBentryALTinterwordstretchfactor}{4}
\providecommand{\BIBentryALTinterwordspacing}{\spaceskip=\fontdimen2\font plus
\BIBentryALTinterwordstretchfactor\fontdimen3\font minus
  \fontdimen4\font\relax}
\providecommand{\BIBforeignlanguage}[2]{{%
\expandafter\ifx\csname l@#1\endcsname\relax
\typeout{** WARNING: IEEEtran.bst: No hyphenation pattern has been}%
\typeout{** loaded for the language `#1'. Using the pattern for}%
\typeout{** the default language instead.}%
\else
\language=\csname l@#1\endcsname
\fi
#2}}
\providecommand{\BIBdecl}{\relax}
\BIBdecl

\bibitem{6484896}
J.~{Brady}, N.~{Behdad}, and A.~M. {Sayeed}, ``Beamspace mimo for
  millimeter-wave communications: System architecture, modeling, analysis, and
  measurements,'' \emph{{IEEE} Trans. Antennas Propag.}, vol.~61, no.~7, pp.
  3814--3827, 2013.

\bibitem{8371237}
I.~{Ahmed}, H.~{Khammari}, A.~{Shahid}, A.~{Musa}, K.~S. {Kim}, E.~{De
  Poorter}, and I.~{Moerman}, ``A survey on hybrid beamforming techniques in
  5g: Architecture and system model perspectives,'' \emph{{IEEE} Commun.
  Surveys Tuts.}, vol.~20, no.~4, pp. 3060--3097, 2018.

\bibitem{8964409}
M.~A.~L. {Sarker}, M.~F. {Kader}, and D.~S. {Han}, ``Rate-loss mitigation for a
  millimeter-wave beamspace mimo lens antenna array system using a hybrid beam
  selection scheme,'' \emph{{IEEE} Syst. J.}, vol.~14, no.~3, pp. 3582--3585,
  2020.

\bibitem{1369651}
D.~J. {Love} and R.~W. {Heath}, ``Limited feedback unitary precoding for
  orthogonal space-time block codes,'' \emph{{IEEE} Trans. Signal Process.},
  vol.~53, no.~1, pp. 64--73, 2005.

\bibitem{7400949}
R.~W. {Heath}, N.~{González-Prelcic}, S.~{Rangan}, W.~{Roh}, and A.~M.
  {Sayeed}, ``An overview of signal processing techniques for millimeter wave
  mimo systems,'' \emph{{IEEE} J. Sel. Topics Signal Process.}, vol.~10, no.~3,
  pp. 436--453, 2016.

\bibitem{8777168}
X.~{Gao}, L.~{Dai}, S.~{Zhou}, A.~M. {Sayeed}, and L.~{Hanzo}, ``Wideband
  beamspace channel estimation for millimeter-wave mimo systems relying on lens
  antenna arrays,'' \emph{{IEEE} Trans. Signal Process.}, vol.~67, no.~18, pp.
  4809--4824, 2019.

\bibitem{8401880}
H.~{Li}, Q.~{Liu}, Z.~{Wang}, and M.~{Li}, ``Transmit antenna selection and
  analog beamforming with low-resolution phase shifters in mmwave miso
  systems,'' \emph{{IEEE} Commun. Lett.}, vol.~22, no.~9, pp. 1878--1881, 2018.

\bibitem{6928432}
L.~{Liang}, W.~{Xu}, and X.~{Dong}, ``Low-complexity hybrid precoding in
  massive multiuser mimo systems,'' \emph{{IEEE} Wireless Commun. Lett.},
  vol.~3, no.~6, pp. 653--656, 2014.

\bibitem{8565897}
K.~{Satyanarayana}, M.~{El-Hajjar}, P.~{Kuo}, A.~{Mourad}, and L.~{Hanzo},
  ``Hybrid beamforming design for full-duplex millimeter wave communication,''
  \emph{{IEEE} Trans. Veh. Technol.}, vol.~68, no.~2, pp. 1394--1404, 2019.

\bibitem{6378483}
H.~{Wang}, Y.~{Li}, X.~{Xia}, and S.~{Liu}, ``Unitary and non-unitary precoders
  for a limited feedback precoded ostbc system,'' \emph{{IEEE} Trans. Veh.
  Technol.}, vol.~62, no.~4, pp. 1646--1654, 2013.

\bibitem{6692283}
Z.~Liu, W.~ur~Rehman, X.~Xu, and X.~Tao, ``Minimize beam squint solutions for
  60ghz millimeter-wave communication system,'' in \emph{2013 IEEE 78th
  Vehicular Technology Conference (VTC Fall)}, 2013, pp. 1--5.

\bibitem{8428615}
M.~A.~L. {Sarker}, M.~F. {Kader}, M.~H. {Lee}, and D.~S. {Han},
  ``Distortion-free golden-hadamard codebook design for miso systems,''
  \emph{{IEEE} Commun. Lett.}, vol.~22, no.~10, pp. 2152--2155, 2018.

\bibitem{Olsen2006}
S.~Olsen, \emph{The Golden Section}.\hskip 1em plus 0.5em minus 0.4em\relax New
  York, NY, USA: Bloomsbury, 2006.

\bibitem{730453}
S.~M. {Alamouti}, ``A simple transmit diversity technique for wireless
  communications,'' \emph{{IEEE} J. Sel. Areas Commun.}, vol.~16, no.~8, pp.
  1451--1458, 1998.

\bibitem{6468994}
S.~{Kundu}, D.~A. {Pados}, W.~{Su}, and R.~{Grover}, ``Toward a preferred 4 x 4
  space-time block code: A performance-versus-complexity sweet spot with
  linear-filter decoding,'' \emph{{IEEE} Trans. Commun.}, vol.~61, no.~5, pp.
  1847--1855, 2013.

\bibitem{Proakis2007}
J.~G. Proakis, \emph{Digital Communications}.\hskip 1em plus 0.5em minus
  0.4em\relax 4th ed. New York, NY, USA: McGraw-Hill, 2007.

\end{thebibliography}

  \end{document}